\documentstyle[epsfig]{aa}

\begin{document}

\thesaurus{12(12.03.4)}

\title{Models of Universe with a Delayed Big-Bang singularity}
\subtitle{III. Solving the horizon problem for an off-center observer}

\author{Marie-No\"elle C\'EL\'ERIER}

\institute{D\'epartement d'Astrophysique Relativiste et de 
Cosmologie, Observatoire de Paris-Meudon, \\
5 place Jules Janssen, 92195 Meudon C\'edex, France}

\mail{Marie-Noelle.Celerier@obspm.fr}

\date{Received 2000 / Accepted}

\maketitle
\markboth{M.N. C\'el\'erier: DBB universe: Solving the horizon problem 
for an off-center observer}
{M.N. C\'el\'erier: DBB universe: Solving the horizon problem for an 
off-center observer}

\begin{abstract}
This paper is the third of a series dedicated to the study of the 
Delayed Big-Bang (DBB) class of inhomogeneous cosmological models of 
Lema\^itre-Tolman-Bondi type. In the first work, it was shown that 
the geometrical properties of the DBB model are such that the horizon 
problem can be solved, without need for any inflationary phase, for an 
observer situated sufficiently near the symmetry center of the model 
to justify the ``centered earth'' approximation. In the second work, we 
studied, in a peculiar subclass of the DBB models, the extent to which 
the values of the 
dipole and quadrupole moments measured in the cosmic microwave background 
radiation (CMBR) temperature anisotropies can support a cosmological 
origin. This implies a 
relation between the location of the observer in the universe and the 
model parameter value: the farther the observer from the symmetry 
center, the closer our current universe to a local homogeneous pattern. 
However, in this case, the centered earth approximation is no longer 
valid and the results of the first work do not apply. We show here that 
the horizon problem can be solved, in the 
DBB model, also for an off-center observer, which improves the consistency 
of this model regarding the assumption of a CMBR large scale anisotropy 
cosmological origin .\\

\end{abstract}

\bigskip
\section{Introduction}

The large scale homogeneity of the universe has been 
recently questioned in an increasing number of works. For instance, the 
controversy over whether the universe is smooth on large scale or 
presents an unbounded fractal hierarchy is not yet ended, and its final 
resolution requires the 
next generation of galaxy catalogs (Martinez, 1999). 
From another point of view, a direct test of the Cosmological Principle 
on our past light cone, up to redshifts approaching unity, has 
been recently proposed, using type Ia supernovae as standard candles 
(C\'el\'erier, 2000). If such tests were to exclude the 
universe homogeneity assumption up to such large scales, and even 
beyond, we would need alternative inhomogeneous models to 
describe its evolution. \\

This article is the third in a series dedicated to the study of a new 
cosmological application of the inhomogeneous Lema\^itre-Tolman-Bondi 
(Lema\^itre, 1933; Tolman, 1934; Bondi, 1947) spherically symmetrical 
dust models. \\

In a first work (C\'el\'erier \& Schneider, 1998, hereafter 
refered to as CS), a subclass of these 
models which solves the standard horizon problem without need for any 
inflationary phase has been identified. This subclass exhibits spatial 
flatness and a conic Big-Bang singularity of ``delayed'' type. 
In this preliminary approach, the observer has been assumed located 
sufficiently near the symmetry center of the model as to 
justify the ``centered earth'' 
approximation. The horizon problem has thus been solved using the 
properties of the null-geodesics issued from the last-scattering 
surface and propagating in a matter dominated region of the universe, 
as seen from a centered observer. \\

However, we stressed in CS a potential difficulty, 
namely the observer at the center. Although such a location 
is not forbidden by scientific principles, it does not account for the 
observed large scale temperature anisotropies of the cosmic microwave 
background radiation (CMBR). \\

The dipole moment in the CMBR anisotropy 
is usually considered to result from a Doppler effect produced by our 
motion with respect to the CMBR rest-frame (Partridge,1988). In the 
second work of the series (Schneider \& C\'el\'erier, 1999, hereafter 
refered to as SC), we assumed that the measured CMBR dipole and 
quadrupole moments can have, totally or partially, a cosmological origin, 
and we studied to which extent they can be reproduced, in a peculiar 
example of our Delayed Big-Bang (DBB) class of models, with no {\it a priori} 
assumption of the observer's location. We have shown that this implies a 
relation between this location and the model parameter 
value, namely the increasing rate $b$ of the Big-Bang function. 
The farther the observer is from the symmetry center, the smaller is the 
value of $b$. \\

The purpose of this present work is therefore to prove that the geometry 
of the DBB model is such that the horizon problem can be 
solved, in principle, with no {\it a priori} constraint on the location 
of the observer. By ``in principle'', we 
mean: provided the null-geodesics are not too distorted in 
the radiation dominated area, such as to prevent them from reconnecting 
before the Big-Bang surface. Were this not the case, it would yield a 
constraint on the model parameters, as discussed in CS. 
This allows us to generalize the previous results of CS to a model with 
an off-center observer, thus improving  the consistency of the assumption 
of a possible cosmological origin of the large scale features of the CMBR 
temperature anisotropies. \\

The above property will have however to be interpreted with care, as 
regards the Ehlers-Geren-Sachs (1968) and almost Ehlers-Geren-Sachs 
(Stoegger et al. 1995) theorems, which are usually considered as robust 
supports for the Cosmological Principle. As was stressed in the 
preceding works, this Principle is not mandatory and does not apply to 
the DBB model. This issue is discussed below in Sect.4. \\

The present paper is organised as follows: a brief reminder of the 
characteristics of the DBB model is given in Sect.2. Arguments and proofs 
are developed in Sect.3. The discussion and conclusion appear in Sect.4.

\section{The inhomogeneous Delayed Big-Bang model}

The main features and properties of the model are here briefly 
mentioned. For a more detailed presentation and a discussion of 
the assumptions retained, the reader is referred to CS and SC. \\

The proposed DBB model is a subclass of the LTB flat model. Its 
line-element, in comoving 
coordinates ($r,\theta,\varphi$) and proper time $t$, is

\begin{equation}
ds^2 = c^2 dt^2 - R'^2(r,t)dr^2 - R^2 (r,t)(d\theta^2 + \sin^2 \theta
d \varphi^2) .  \label{eq:1}
\end{equation}
An appropriate choice of the radial coordinate $r$ yields
\begin{equation}
R(r,t)=\left({9GM_0\over 2}\right)^{1/3} r[t-t_0(r)]^{2/3} . 
\label{eq:2}
\end{equation}

The Big-Bang function, $t_0(r)$, verifies \footnote {An interesting 
example, studied in CS and SC, is the subclass with $t_0(r)=br^n$, 
$b>0$, $n>0$.}

\begin{eqnarray}
t_0(r=0) &=& 0 , \nonumber \\
t'_0(r) &>& 0  \qquad  \hbox{for all} \qquad r , \nonumber\\
5t'_0(r) + 2rt"_0(r) &>& 0 \qquad \hbox{for all} \qquad r , 
\nonumber \\
rt'_0|_{r=0} &=& 0  . \label{eq:3}\\
\nonumber\end{eqnarray}

The physical singularity of the model - i.e. the first surface, 
encountered on a backward path from  ``now'', where the energy 
density and the invariant scalar curvature go to infinity - is the 
shell-crossing surface, represented in the $(r,t)$ plane by the 
curve:

\begin{equation}
3t-3t_0(r)-2rt'_0(r)=0 . \label{eq:4}
\end{equation}

As the energy density increases approaching 
this surface, radiation becomes the dominant component in the 
universe, pressure can no longer be neglected, and the LTB model 
no longer holds. The region between the Big-Bang surface 
$t=t_0(r)$ and the shell-crossing one is thus excluded 
from the part of the model retained to describe the matter dominated 
region of the universe, which we discuss here. \\  

The optical depth of the universe to Thomson scattering is approximated 
by a step function (see SC). The last-scattering 
surface is thus defined, in the local thermodynamical equilibrium 
approximation (see CS), by its temperature, $T=4 000$ $K$, as is  
the now-surface, $T=2.73$ $K$, where the observer is located. The equal 
temperature surfaces verify

\begin{equation}
t=t_0(r)+{r\over 3}t'_0(r)+{1\over 3} \sqrt{r^2t'^2_0(r)+
{3S(r)\over 2\pi G k_B a_n m_b T^3}} .  \label{eq:5}
\end{equation}

The value of the entropy function $S(r)$ is assumed to 
be constant with $r$. The shell-crossing surface is thus asymptotic to
every (monotonically increasing with $r$) $T=const.$ curve. \\

An observer located at a distance from the center sees an 
axially symmetrical universe in the center direction. In the geometrical 
optics approximation, the light 
travelling from the last-scattering surface to this observer follows
null-geodesics, which is thus legitimate to consider in the 
meridional plane. The photon path is uniquely defined
by the observer's position $(r_p,t_p)$ and the angle $\alpha$ between
the direction from which the light ray comes 
and the direction of the center of the universe, $C$.\\

The microwave radiation observed in the direction of this symmetry center 
follows a light-cone issued from point $D$ on 
the last-scattering surface, then passes through the center 
$(r=0)$ and reaches the earth at point $O$. Observed in the opposite 
direction, it starts from point $E$, and travels on the $EO$ 
null-geodesic to the observer (see Fig.2). \\

The radial null-geodesic equation, as established in CS, is

\begin{equation}
{dt\over dr}=\pm {1\over 3c}\left({9GM_0\over 2}\right)^{1/3}{
3t-3t_0(r)-2rt'_0(r)\over [t-t_0(r)]^{1\over 3}} . \label{eq:22}
\end{equation}

It is easy to see that, with a function $t_0(r)$ verifying 
conditions (\ref{eq:3}), for a fixed $t$, $\left| dt\over dr\right|$ decreases 
with increasing $r$, and thus: $r_D<r_E$. \\

If observed with an angle $\alpha$ in the inward direction, a 
light beam issued from  point $A$ on the last-scattering surface 
approaches $C$ to a comoving distance $r_{min}$, then proceeds 
toward $O$. In the outward (opposite) direction, it follows the 
$BO$ geodesic (see Figs.1 and 2). \\

The corresponding null-geodesics are solutions of the system of 
differential equations established in SC (in units $c=1$):

\begin{equation}
{dt\over d\lambda}=k^t , \label{eq:19}
\end{equation}
\begin{equation}
{dr\over d\lambda}=\pm {1\over R'} 
\left[(k^t)^2 -\left(R_p \sin \alpha \over R \right)^2
\right]^{1/2} , \label{eq:20}
\end{equation}
\begin{equation}
{dk^t\over d\lambda}=-{\dot{R}'\over R'}(k^t)^2 +
\left({\dot{R}'\over R'}- {\dot{R}\over R} \right) \left(
R_p \sin \alpha \over R \right)^2 . 
\label{eq:21}
\end{equation}
with a plus sign in 
Eq.(\ref{eq:20}) from $O$ to $r_{min}$ (${dr\over d\lambda}>0$), 
and a minus sign from $r_{min}$ to $A$ (${dr\over d\lambda}<0$). The equations 
corresponding to the observer looking outward (OB curve) require a minus 
sign.\\

If one considers Eq.(\ref{eq:20}), for a fixed $t$ 
and for a same variation $d\lambda$ of the $\lambda$ affine parameter, 
the absolute value of the radial coordinate variation $dr$ is smaller 
with $\alpha\neq 0$ or $\pi$ than with $\alpha=0$ or $\pi$. It follows 
that $r_D<r_A<r_B<r_E$. \\

With $\alpha$ taking every value between $-{\pi \over 2}$ and 
$\pi \over 2$, and considering the two oppposite directions of 
sight, the CMBR is observed from $O$ as coming from a set of points 
each located on a 2-sphere belonging to the the subset $\lbrace r=const., 
r_D<r<r_E \rbrace$ on the last-scattering surface $T=4 000 K$. \\

To prove that this set of points can be causally connected, it is 
sufficient to show that there is at least one forward radial light-cone, 
i.e. issued from a point $(r=0, t>0)$ including the $DE$ subset.

\includegraphics[height=8cm,width=8cm,angle=-90]{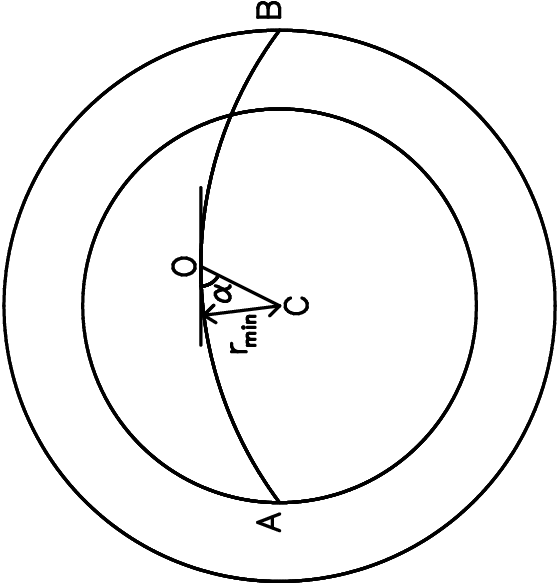}
\begin{center}
{\bf Fig.1 : The CMBR observed from $O$ with an angle 
$\alpha$.} Schematic illustration of the trajectory of two CMBR light beams 
received by the observer $O$ and making an angle $\alpha$ with the 
direction of the symmetry center $C$ of the universe. The inner 
circle is the 2-sphere on the last-scattering surface from which 
the beam issued from point $A$ is emitted (the observer looks inward). 
The outer circle is the one from which the beam 
starting from $B$ is emitted (the observer looks outward).
\end{center}

\includegraphics[height=8cm,width=8cm,angle=-90]{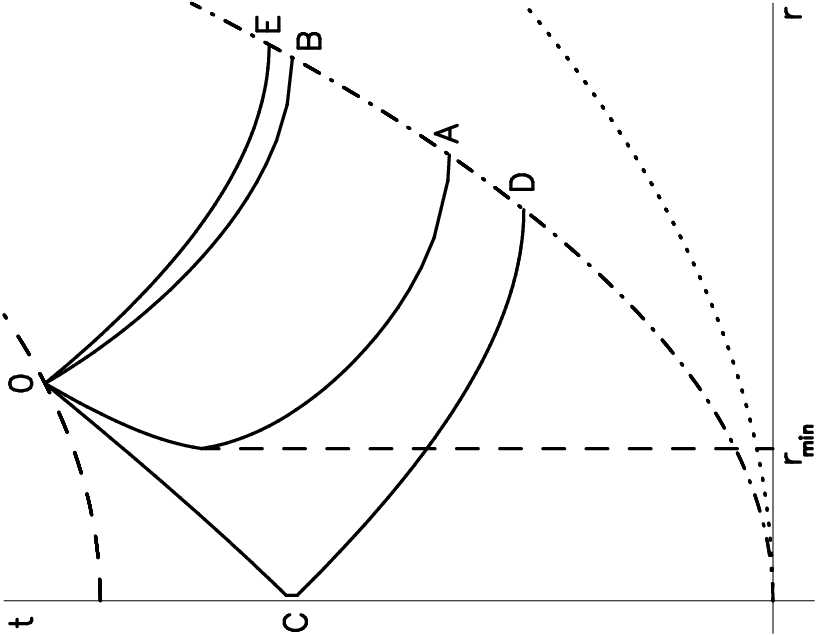}
\begin{center}
{\bf Fig.2 : The CMBR observed from $O$ in the $(r,t)$ plane.}
The dotted curve represents the Big-Bang surface $t=t_0(r)$. The broken  
line with dots represents the shell-crossing singularity  and the 
last-scattering surfaces that cannot be resolved at the scale of 
the figure. The broken curve is the now-surface $T=2.73 K$. 
The solid lines are the light-cones.
\end{center}

\section{Solving the horizon problem}

An inspection of Eqs.(\ref{eq:4}) and (\ref{eq:22}), as done in CS, shows 
that the shell-crossing singularity surface is null: it cannot be crossed 
by any ingoing null geodesic. \\

The solution of the horizon problem, for an off-center observer in a DBB 
model, can thus proceed from its representation using a Penrose-Carter 
conformal diagram (see Fig.3) \footnote {Only half of the 
diagram is drawn, as permitted by spherical symmetry.}. \\

\includegraphics[height=6cm,width=10cm,angle=-90]{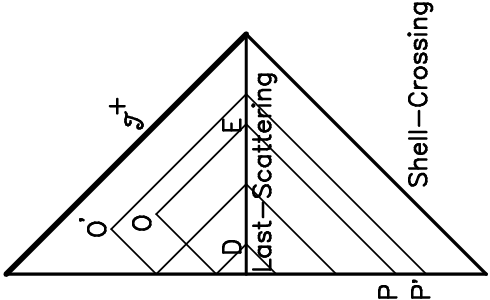}
\begin{center}
{\bf Fig.3. : Penrose-Carter diagram sketching the permanent solution 
of the horizon problem by the DBB model.} The current observer 
$O$ sees, on the last-scattering surface, a causally connected (DE)
region, included in the forward light cone of $P$. The same will hold when 
the $O'$ point is reached, and from any other time in the observer's past 
or future history. \\
\end{center}

The light-like character of the shell-crossing surface forces all matter to 
be causally connected at $t=0$, and any finite region to have been in causal 
contact at some $t>0$. \\

The horizon problem is thus solved permanently in this model, {\it a priori}, 
for any location of the observer \footnote {As pointed out in Sect.4 (below), 
constraints yielded by observational data can be imposed upon the observer 
location, but solving the horizon problem, in the dust approximation, 
does not provide any constraint of this kind.}. \\

It is worth emphasizing that if the inflationary assumption also 
solves the horizon problem, it does so only temporarily. In effect, if one 
considers 
the horizon problem in a standard FLRW universe, as sketched in Fig.4, 
the Big-Bang surface is space-like. It thus implies the existence of a 
limiting point $L$, in the history of the observer $O$, beyond which 
the observer sees, on the last-scattering surface, some no causally 
connected points. The current observer, being located above 
$L$, is confronted with this horizon problem. \\

\includegraphics[height=6cm,width=8.4cm,angle=-90]{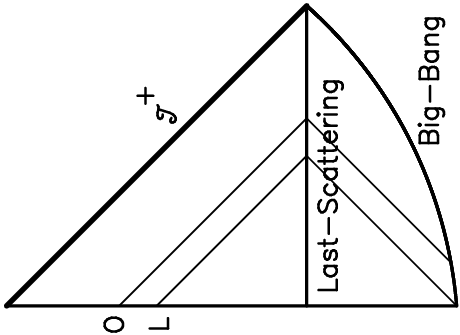}
\begin{center}
{\bf Fig.4. : Penrose-Carter diagram showing the horizon problem in a 
FLRW universe.} The thin lines represent the light-cones. The CMBR, as 
seen by the observer located on the vertical axis, corresponds to the 
intersecting point of the observer backward light-cone and the 
last-scattering line. For a complete causal connection between every point 
seen in the CMBR, the backward light-cone issued from this intersecting 
point must reach the vertical axis before the Big-Bang curve. $L$ is thus 
the limiting time beyond which the observer $O$ experiences the horizon 
problem. 
\end{center}

The solution proposed by the inflationary assumption is presented in 
Fig.5. Adding an inflationary phase in the early history of the universe 
amounts to adding a slice of de Sitter space-time betwen the Big-Bang and 
the last-scattering, thus postponing the limit $L$ when the observer can 
see non-causally connected points in the CMBR. Inflation thus 
solves the horizon problem, but only temporarily. \\

\includegraphics[height=6cm,width=9.7cm,angle=-90]{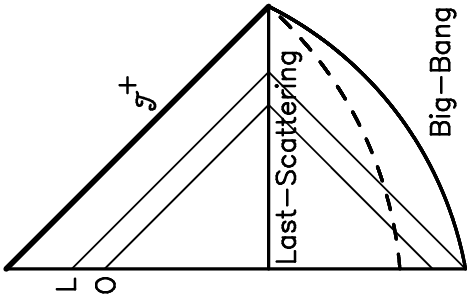}
\begin{center}
{\bf Fig.5. : Penrose-Carter diagram showing the temporary solution of 
the horizon problem by inflation.} The slice of de Sitter space-time 
corresponding to an inflationary phase, and 
added to Fig.4, is shown between the Big-Bang and dashed lines. 
$L$ is thus postponed, allowing the current observer $O$ to see a 
causally connected CMBR. When time 
elapses and the observer reaches the above $L$ region, the horizon problem 
returns.
\end{center}

\section{Discussion and conclusion}

In CS we identified a subclass of the LTB models with a 
Big-Bang of ``delayed'' type which solves the standard horizon problem 
without need for any inflationary phase. 
In this preliminary approach, the observer was assumed located 
sufficiently near the symmetry center of the model as to justify 
the ``centered earth'' approximation. \\

Here, we report a further analysis of the properties of the DBB 
model to show that this model solves the 
horizon problem even with an off-center observer. The model is 
thus relieved of a prescription that could be considered as 
``unnatural''. \\

The model is also provided with a new free parameter, the spatial location 
$r_p$ of the earth in the universe, which accounts for the large 
scale inhomogeneities observed in the CMBR temperature anisotropies. 
The measured dipole and quadrupole moments of these anisotropies set 
bounds on the correlated values of this $r_p$ parameter and of the 
local deviation of the model from homogeneity, accounted for by the slope 
of the Big-Bang function. A possible cosmological part of these large 
scale features seen in the CMBR, if once observationally identified, 
would select an even narrower curve in the parameter space of the model, 
as shown in SC. \\

It is of the utmost importance to stress that, as was the case with 
a centered observer, these results hold for any universe arbitrarily 
{\it locally} close to the FLRW $t_0(r)=const.$ asymptotic model. The 
only requirements to be fulfilled are conditions (\ref{eq:3}) which 
are obviously compatible with an almost ``flatness'' of the 
Big-Bang function up to comoving shells arbitrarily far out the $r_p$ 
shell where the observer is located. The properties of the light-cones 
are preserved as long as this function does not reduce to a mere 
constant. For instance, the subclass retained in SC, with $t_0(r)=br$, 
reduces to a FLRW model for $b$ equal to zero, but fulfills the 
conditions (\ref{eq:3}) for $b$ as small as one wishes, provided $b$ does 
not vanish. No bound can therefore be {\it a priori} inferred on 
the observer location, as, according to SC, an arbitrarily small 
value of $b$ corresponds to an arbitrarily large value of $r_p$, and 
conversely
\footnote {Such bounds could in fact proceed from further analyses of the 
model in the light of other theoretical consideratioins or observational 
data, but solving the horizon 
problem does not yield any constraint of this kind.}. \\

A point worth discussing here is the validity of this claim 
as regards the almost Ehlers-Geren-Sachs (AEGS) theorem (Stoegger et al. 
1995). This theorem states that ``if all 
fundamental observers measure the cosmic background radiation to be 
almost isotropic in an expanding universe region, then that univere is 
locally almost spatially homogeneous and isotropic in that region.'' 
The U region considered by the AEGS authors is ``the region within and 
near our past light cone from decoupling to the present day''. It is 
easy to see that small $b$ DBB models fulfills the AEGS prescription, 
as they can remain ``close'' to FLRW models for shells located between 
the center and an arbitrarily large value of the comoving radial 
coordinate, including the AEGS region. But the further away part of 
these small $b$ DBB models infinitely diverges from homogeneity. On 
the contrary, the AEGS theorem does not apply to large $b$ DBB models, 
implying an observer close to the center. The founding assumption of 
this theorem, namely the local Copernican principle applied to the 
U region, is not retained in this case. As was discussed in CS, 
such a choice is perfectly compatible with all available observational 
data and scientifically grounded principles.\\

It is also interesting to note that, contrary to the inflationary 
assumption which restores causality between the different points seen 
in the CMBR, but only temporarily, the DBB model provides a permanent 
solution to the horizon problem, whatever the position of the observer 
on his world line. \\

In the prospect of future observational tests of the large scale 
(in)homogeneity of the universe, the development of other interesting 
inhomogeneous models must be regarded as an 
important issue. However, this presented result is only a first 
improvement in the release of the simplifying assumptions (retained in CS) 
for a preliminary study of the properties induced by a ``delayed 
Big-Bang''. Other analyses are still needed, among which the release 
of the spatial 
spherical symmetry of the model and of the dust approximation should be 
considered as priorities. \\

{\it Aknowledgements.} The author thanks Brandon Carter for a fruitful 
discussion leading to the use of Penrose-Carter diagrams, as an 
enlightening tool.

\end{document}